# Jirim kuark - mikroskopi elektron attoskala ke atas proton*

Wan Ahmad Tajuddin Wan Abdullah
*Kumpulan Penyelidikan Zarah Keunsuran*
*Jabatan Fizik, Universiti Malaya, 50603 Kuala Lumpur*

Kita imbas eksperimen perlanggaran elektron(/positron) ke atas proton pada tenaga tinggi di HERA, dengan memberi fokus ke atas eksperimen ZEUS. Kita perihalkan pengesan ZEUS itu, termasuk sistem pemerolehan datanya, dan melihat suatu bahagian khusus, iaitu sistem elektronik kawalan bacaan keluar kalorimeternya, secara lebih terperinci. Cara analisis data diperihalkan, dan hasil-hasil berkenaan struktur proton diberikan. Didapati model kuark naif tidak menerangkan hasil yang diperolehi pada pecahan momentum kecil, dan ini memerlukan kromodinamik kuantum. Juga didapati bahawa tindakbalas belauan, dengan jurang kecepatan yang besar, menyumbang dengan banyaknya; usaha sedang dijalankan untuk memahaminya.

**Prasembang**

Jemputan bagi saya membentangkan kertas plenari di Pulau Duyong ini sangat bererti bagi saya kerana di sini lebih-kurang satu kurun yang lalu hidupnya Tok Syeikh Duyong atau Wan Abdullah bin Wan Muhammad Amin, yang merupakan seorang ulama yang ulung. Pulau Duyong telah merupakan pusat ilmu sehingga disebut "Lidah Terengganu", mungkin juga kerana bentuknya di muara Sungai Terengganu, dan dikatakan Sultan Terengganu waktu itu sendiri pergi ke Pulau Duyong untuk mempelajari ilmu. Satu peninggalan ialah Kota Lama Duyong, yang dibina oleh keturunan beliau, yang masih berdiri hanya lebih-kurang seratus meter dari tempat persidangan ini.

Yang membawa keertian besar ialah saya merupakan Wan Ahmad Tajuddin bin Wan Abdullah bin Wan Sulaiman bin Wan Daud bin Tok Syeikh Duyong.

**Pengenalan kepada struktur proton**

Model kuark-parton pada tahun-tahun 1960an menyatakan bahawa hadron-hadron adalah terdiri daripada kuark; 3 kuark dalam barion, dan pasangan kuark-antikuark dalam meson. Substruktur proton telah ditunjukkan oleh eksperimen perlanggaran elektron yang dipecut ke atas proton rehat di pemecut linear SLAC di Stanford, serupa seperti

---
* Kertas plenari jemputan di Persidangan Fizik Kebangsaan (PERFIK) 2007, Pulau Duyong, Kuala Terengganu, Malaysia, 26-28 Disember 2007

Rutherford menunjukkan ada substruktur dalam atom menerusi perlanggaran zarah alfa ke atas atom aurum.

Daripada 6 perisa kuark yang ada, dua yang teringan, kuark naik ($u^{+2/3}$) dan kuark turun ($d^{-1/3}$) membina hadron-hadron 'stabil' seperti proton (uud) dan neutron (udd), sementara yang lain mereput kepada yang dua ini. Kuark diikat dalam hadron akibat cas daya nukleus kuat, yang menurut kromodinamik kuantum, terdiri daripada 3 jenis atau 'warna' ('merah', 'hijau' 'biru'), dan diantarakan oleh boson gluon, yang juga membawa warna (dan antiwarna). Hanya objek-objek tak berwarna (atau 'putih'), iaitu merah-hijau-biru dalam barion dan merah-antimerah, dll dalam meson. Objek berwarna seperti kuark atau gluon tunggal, menyebabkan pengeluaran pasangan warna dari vakum, dan akhirnya menghasilkan 'jet' hadron menerusi proses yang disebut serpihan. Teori medan kuantum yang menerangkan salingtindak berwarna ini ialah Kromodinamik Kuantum (QCD).

**HERA dan ZEUS**

Pemecut HERA di DESY, Hamburg, Jerman, memecut elektron (atau positron) sehingga 27.6 GeV, dan proton dalam arah bertentangan, sehingga 920 GeV. Perlanggaran berlaku pada beberapa kedudukan, dan dipantau oleh pengesan-pengesan zarah yang terhasil. Eksperimen-eksperimen ini telah mengambil data dari 1992 hingga 2007.

Satu daripada pengesan di HERA ialah pengesan ZEUS. Pakatan ZEUS ini terdiri daripada lebih-kurang 450 saintis dari lebih-kurang 50 institusi dari lebih-kurang 15 negara. Kumpulan kami dari Universiti Malaya telah diterima bersama pakatan ZEUS semenjak 2005.

**Pengesan ZEUS**

HERA telah menyampaikan lebih-kurang 530 $pb^{-1}$ luminositi bim keseluruhannya. Bagi ZEUS, ini bermakna lebih-kurang 350 juta peristiwa dan lebih-kurang 80 TB data mentah. Pengesan ZEUS (lihat rujukan [1] untuk pemerihalan lebih sempurna) merupakan pengesan serbaguna dengan pengkalorimeteran elektromagnet dan hadron, jejakan bercas dalam medan magnet solenoidan, dan pengesanan muon sebagai komponen-komponen utama, yang direkabentuk dengan hermetisiti maksimum supaya momentum serenjang hilang akibat penghasilan neutrino dapat dikesan. Gambarajah 1 menunjukkan rajah berskema pengesan ZEUS.

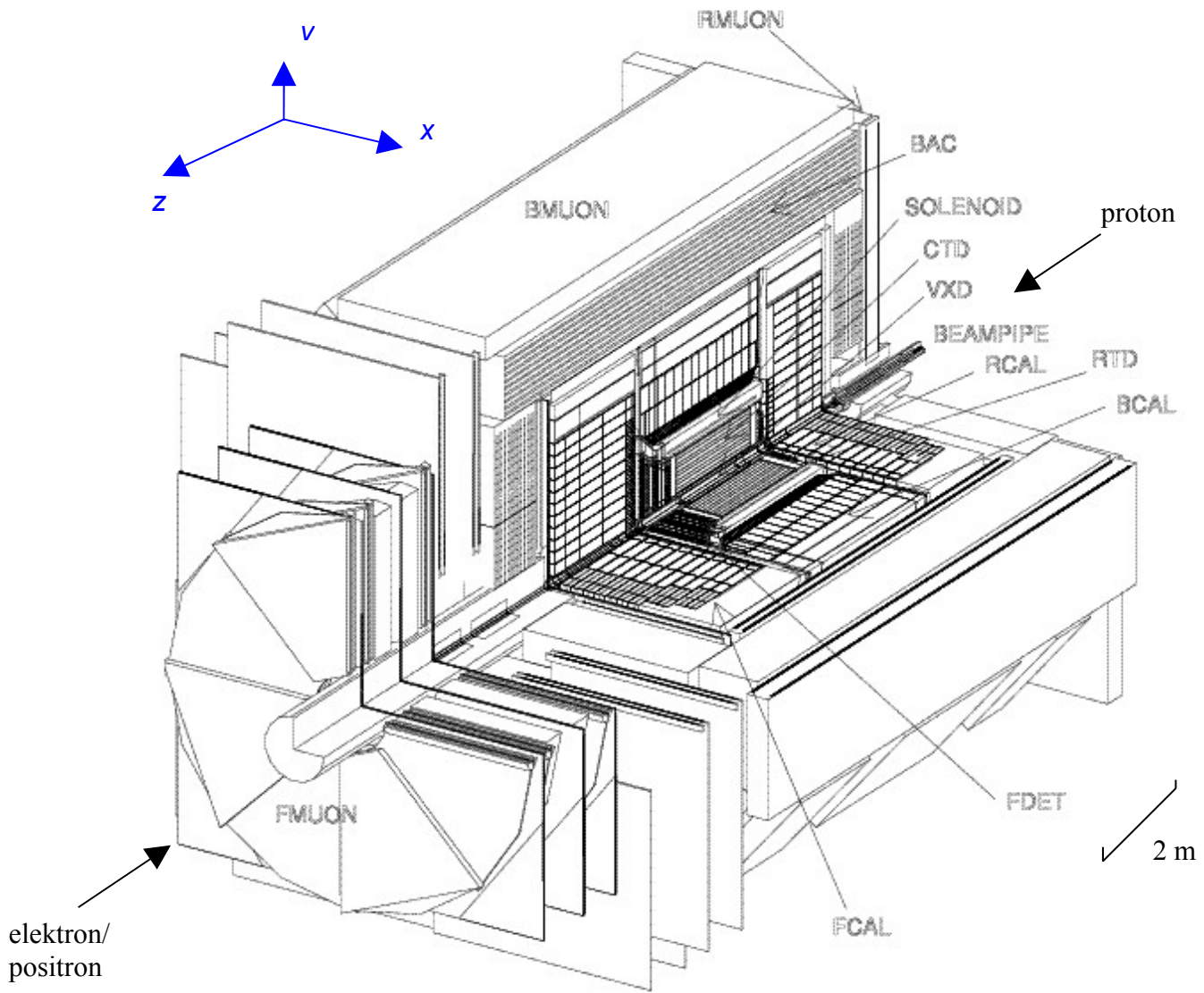

*Gambarajah 1: Skema pengesan ZEUS. Sistem kordinat yang digunakan ditunjukkan. BEAMPIPE: paip bim, VXD: pengesan verteks (jaluran silikon), CTD: pengesan jejak pusat, SOLENOID: magnet solenoidan, BCAL: kalorimeter tong, BAC: kalorimeter sokongan, BMUON: pengesan muon tong, RTD: pengesan jejak belakang, RCAL: kalorimeter belakang, RMUON: pengesan muon belakang, FDET: pengesan hadapan, FCAL: kalorimeter hadapan, FMUON: pengesan muon hadapan.*

## CAL dan ROC

Kalorimeter yang digunakan merupakan kalorimeter menyampel, dibina daripada plat-plat uranium tersusut berselang-seli dengan pengsintilasi plastik sebagai bahan aktif. Nisbah ketebalan dipilih bagi memperolehi pampasan dan peleraian terbaik untuk hadron. Rekabentuk yang digunakan memberikan ukuran tepat tenaga hadron dan jet: $\sigma(E)/E = 0.35/\sqrt{E}$ ($E$ dalam GeV), dengan peleraian sudutan bagi jet lebih baik daripada 10 mrad. Kita boleh bezakan antara hadron dan elektron menerusi pola mendapan tenaga berbeza seperti yang ditunjukkan dalam Gambarajah 2. Ketepatan tenaga bagi elektron diberikan sebagai $\sigma(E)/E = 0.18/\sqrt{E}$. Kalorimeter ini juga mempunyai peleraian masa 1 ns dan menjangkau sudut kutuban dari $\theta = 2.2°$ sehingga $176.5°$.

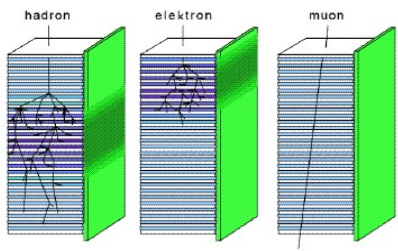

*Gambarajah 2. Pola mendapan tenaga berlainan bagi zarah berlainan dalam kalorimeter.*

Secara keseluruhan, ada 3 bahagian kalorimeter: Kalorimeter Tong atau Barrel Calorimeter (BCAL), Kalorimeter Hadapan atau Forward Calorimeter (FCAL), dan Kalorimeter Belakang atau Rear Calorimeter (RCAL). Setiapnya adalah terdiri daripada modul-modul yang serenjang (BCAL) dan selari (FCAL, RCAL) kepada bim, yang merupakan menara-menara berkeratan rentas 20×20 cm$^2$, dengan bahagian hadapannya bahagian elektromagnetan, dan bahagian belakangnya bahagian hadronan. Bacaan keluar dibuat oleh penganjak jarakgelombang yang diganding kepada tiub fotopengganda di kedua belah, membolehkan pembinaan semula kedudukan juga.

Lintasan bim berlaku setiap 96 ns, menentukan keperluan kadar pembacaan keluar data. Pilihan perlu dibuat samada peristiwa yang berlaku adalah yang menarik, yang perlu dibaca keluar. Bacaan keluar bolehlah pada kadar lebih perlahan apabila diambilkira kadar peristiwa menarik yang lebih rendah daripada kadar lintasan. Data di komponen-komponen pengesan ditalipaipkan untuk memberi masa pengiraan mudah dibuat untuk memutuskan samada peristiwa berkenaan mahu diterima atau tidak. Pencetus 3 tahap digunakan dalam ZEUS, iaitu ada 3 tahap keputusan dalam memutuskan penerimaan sesuatu peristiwa itu: pada tahap komponen yang digabung dengan data masih berbentuk analog (GFLT), yang mengurangkan kadar bacaan daripada 10 Mhz kepada 1 kHz; setelah pendigitan data (GSLT), yang mengurangkan kadar bacaan daripada kHz kepada ratusan Hz, memberi masa untuk pembinaan peristiwa; dan setelah peristiwa dibina (TLT), yang akhirnya memberikan bacaan data dalam kadar beberapa Hz.

Bagi CAL, bacaan mula-mula dimasukkan ke dalam talipaip analog (siri kapasitor) menunggu keputusan GFLT – jika diterima, data seterusnya dihantar kepada litar pendigitan dan seterusnya, jika tidak, talipaip ditulis atasnya dengan data baru tanpa data lama dihantar ke pendigitan. Kawalan proses ini dan juga kawalan larian kalibrasi, Kawalan Baca Keluar (ROC), dilakukan oleh 5 modul seperti dalam Gambarajah 3.

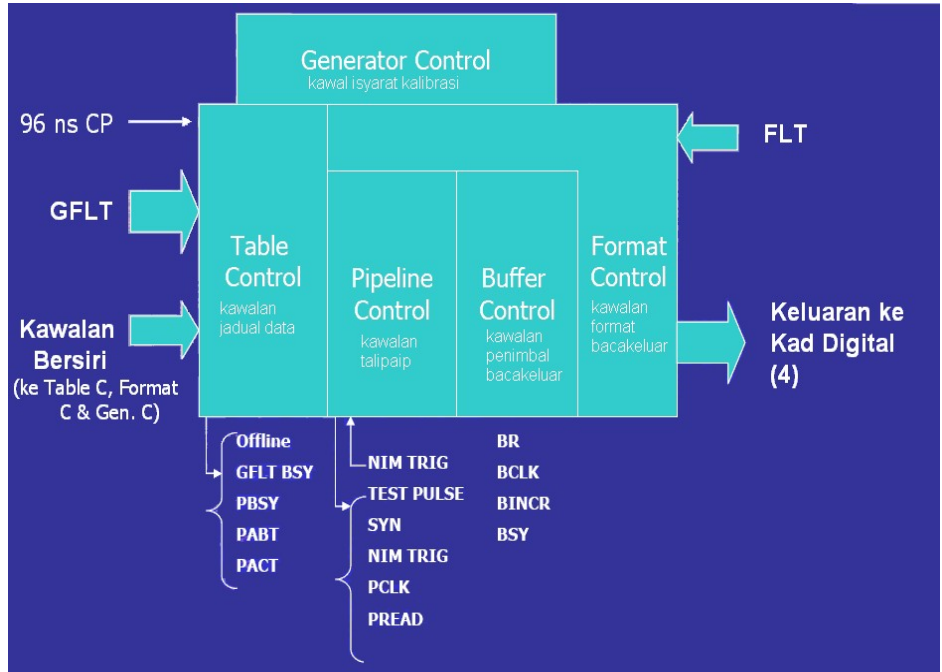

*Gambarajah 3. Kawalan Bacaan Keluar untuk Kalorimeter*

Kumpulan Universiti Malaya telah mengimplementasi semula ROC menggunakan Tatasusunan Get Teraturcarakan Medan (FPGA). Rekabentuk teratas ditunjukkan dalam Gambarajah 4, sementara beberapa hasil simulasi diberikan dalam Gambarajah 5.

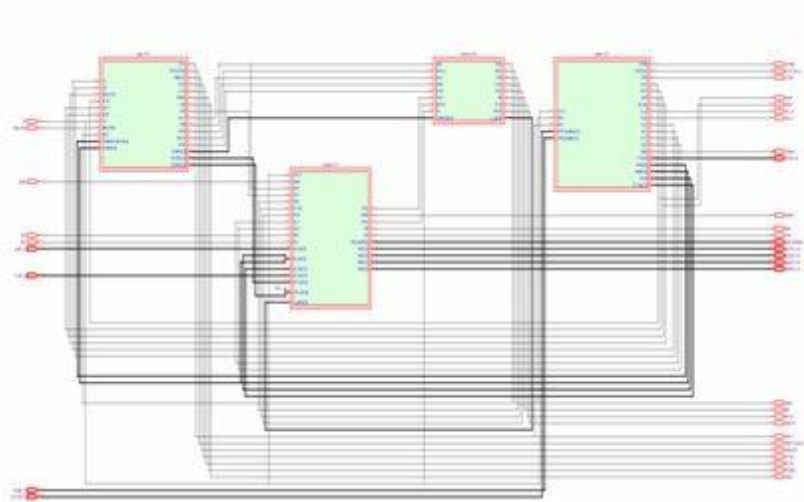

*Gambarajah 4. Rekabentuk FOC dalam FPGA*

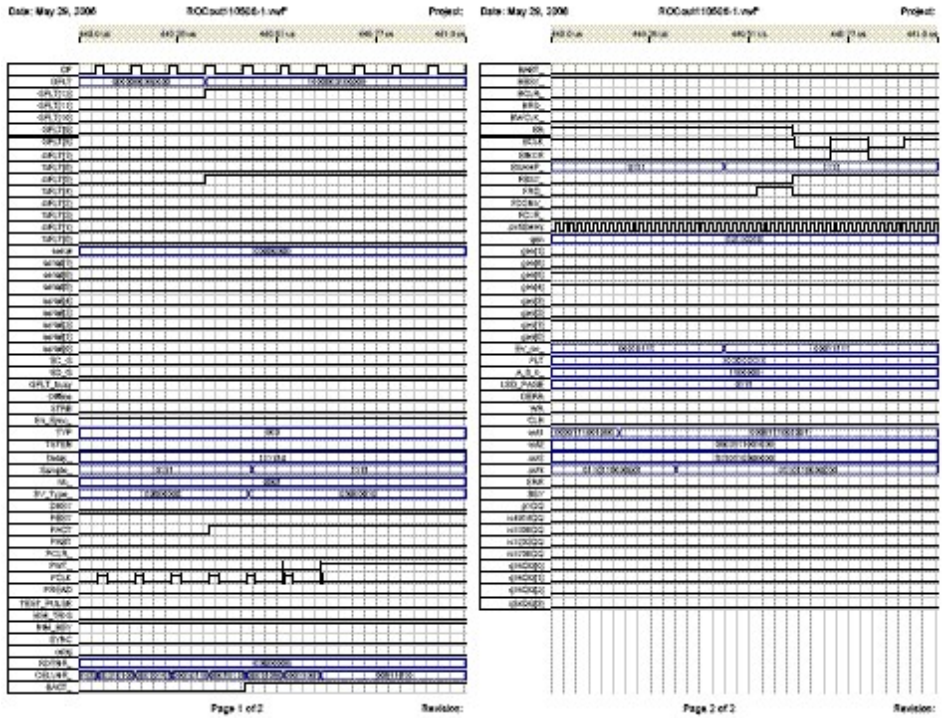

*Gambarajah 5. Simulasi ROC dalam FPGA*

Data yang dibacakeluar kemudiannya dianalisis di luar talian menggunakan kemudahan komputer, termasuklah ladang komputer dengan multiCPU dan grid komputer yang merentasi beberapa benua. Grid ini utamanya digunakan untuk menghasilkan peristiwa buatan secara simulasi Monte Carlo bagi tujuan penentusahan analisis dan pengiraan kecekapan analisis. Perisian untuk paparan visual peristiwa-peristiwa juga wujud.

**DIS**

Sekarang kita lihat fizik yang boleh dikaji dalam perlanggaran elektron/positron-proton. Dalam perbincangan ini kita hanya lihat sebahagian sahaja, terutama yang berkaitan dengan struktur proton.

Suatu proses utama dalam perlanggaran elektron-proton ialah Serakan Dalam Takkenyal (DIS), yang diberikan dalam Gambarajah 6 (untuk hasil-hasil terbaharu dan lebih meluas, sila rujuk [2]). Kemayaan foton tukarganti diberikan sebagai negatif kuasadua momentum-4nya,

$$Q^2 = -q^2$$

dan proses DIS dicerminkan oleh nilai $Q^2$ yang besar (lebih besar daripada jisim proton). Kalau ini dikaitkan dengan tenaga dan seterusnya jarakgelombang foton itu, $Q^2$ yang tinggi mewakili kuasa peleraian yang tinggi. Pecahan momentum kuark (dalam proton) dihentam boleh diterbitkan sebagai

$$x = \frac{Q^2}{2P \cdot q}$$

sementara ketakkenyalan, iaitu pecahan tenaga termendap dalam rangka rehat proton,

$$y = \frac{P \cdot q}{P \cdot k} = \frac{Q^2}{xs}$$

di mana tenaga pusat jisim kuasadua,

$$s = (P+k)^2.$$

Tenaga pusat jisim sistem γp ialah

$$W = (q+p)^2.$$

Dalam proses ini, hanya ada dua pembolehubah bebas, sementara yang lain bergantung kepada yang dua ini.

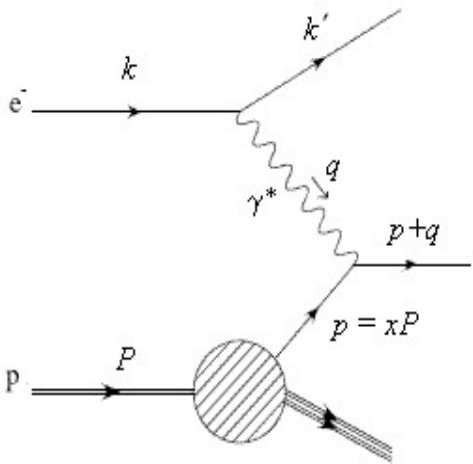

*Gambarajah 6. Serakan Dalam Takkenyal (DIS). Label mewakili momentum-4 yang berkenaan*

Dalam serakan dalam takkenyal, jika foton ditukarganti dari elektron (atau positron) tuju, maka kita ada kes arus neutral (NC); begitu juga dengan tukarganti $Z^0$:

$$e^{\pm}p \rightarrow e^{\pm}X$$

Kita ada arus bercas (CC) dalam kes tukarganti $W^{\pm}$ di mana elektron (positron) tuju berubah menjadi (anti)neutrino:

$$e^- p \to \nu X.$$

Suatu contoh peristiwa CC yang dilihat dalam pengesan ZEUS ditunjukkan dalam Gambarajah 7.

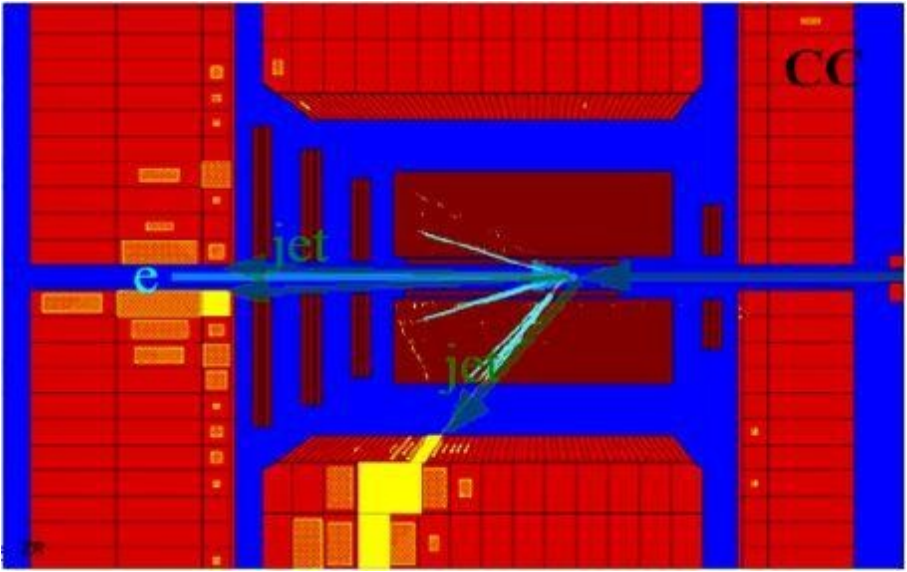

*Gambarajah 7. Peristiwa arus bercas dilihat oleh ZEUS. Ada momentum serenjang yang 'hilang' (tak berimbang) akibat neutrino yang dihasilkan.*

Keratan rentas kebezaan untuk DIS diberikan oleh

$$\frac{d^2\sigma(e^{\pm}p)}{dx\,dQ^2} = \frac{2\pi\alpha^2}{xQ^4}\left[Y_+ F_2(x,Q^2) - y^2 F_L(x,Q^2) \mp Y_- xF_3(x,Q^2)\right](1+\delta_r)$$

di mana

$$Y_{\pm} = 1 \pm (1-y)^2$$

Dalam ungkapan ini, $Y_+$ adalah kesan spin, $F_L$ ialah fungsi struktur membujur, $F_3$ adalah sumbangan elektrolemah, dan $\delta_r$ ialah pembetulan sinaran. Dalam DIS, dengan $y$ dan $x$ rendah, sebutan pertama mendominasi.

Tenaga operasian HERA membolehkan peninjauan ruang fasa $x$-$Q^2$ yang sebelum ini tidak dijangkau oleh eksperimen-eksperimen sasaran tetap, seperti ditunjukkan dalam

Gambarajah 8. Eksperimen-eksperimen HERA dapat melihat julat $Q^2$ yang lebih luas, dan rantau $x$ kecil yang belum dapat dilihat sebelumnya.

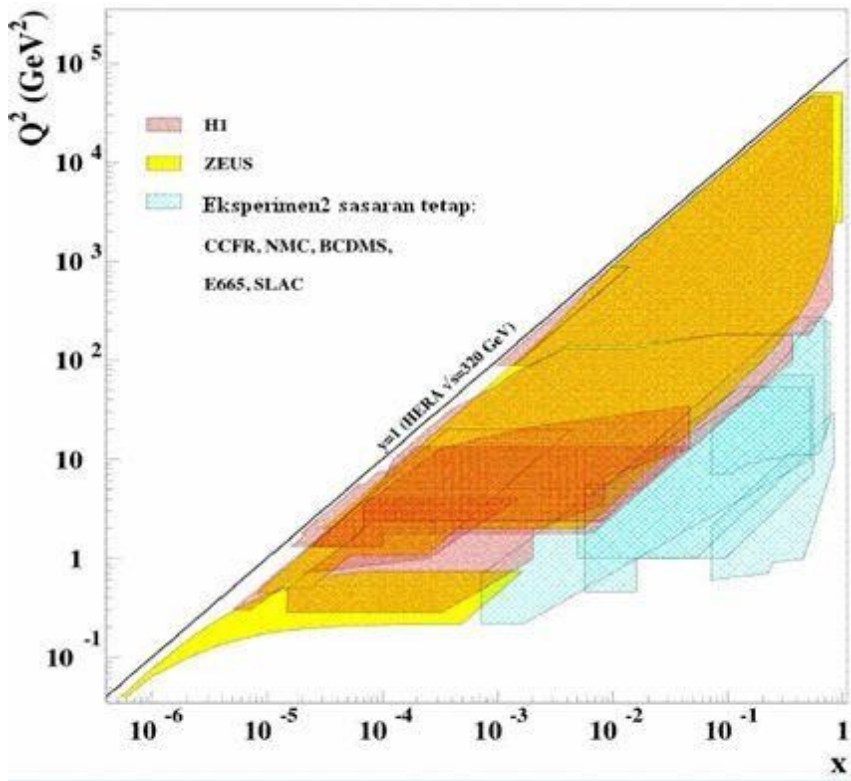

*Gambarajah 8. Ruang $x$-$Q^2$ yang dijangkau eksperimen-eksperimen. Dari [3].*

**Substruktur kuark**

Sudah tentu perkara pertama yang harus dilihat, selari dengan falsafah eksperimen Rutherford, dengan tenaga yang lebih tinggi ini, ialah samada kuark juga mempunyai substruktur.

Satu cara untuk mencari substruktur kuark ialah dengan mengandaikan salingtindak sentuhan eeqq dengan kuark bersaiz terhingga menerusi suatu faktor bentuk. Padanan data, seperti dalam Gambarajah 9, memberikan had tinggi kepada jejari kuark,

$$r_q < 0.85 \times 10^{-18} \text{ m}$$

daripada hasil ZEUS [4].

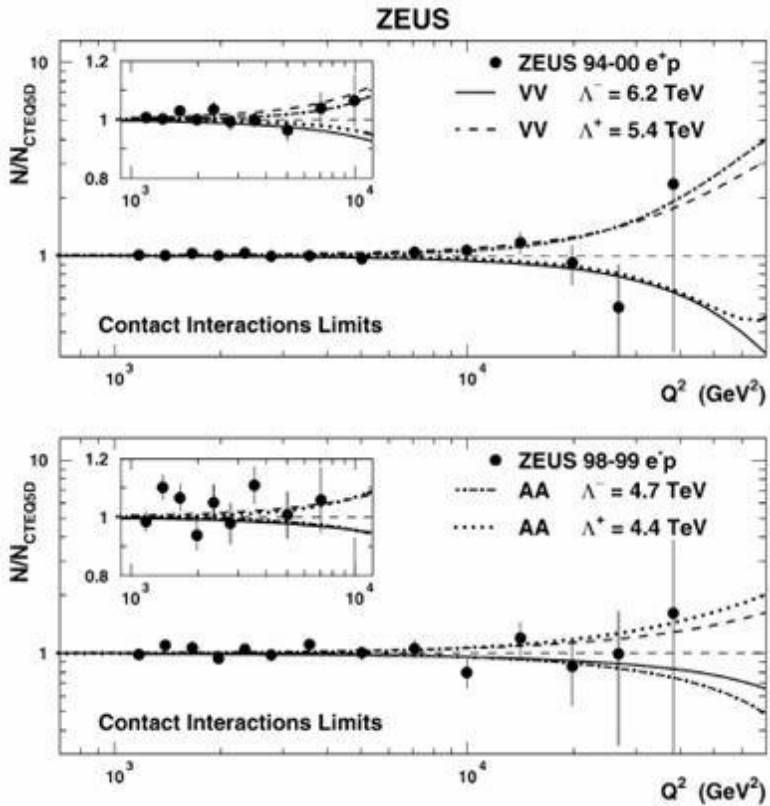

*Gambarajah 9. Had kepada salingtindak sentuhan dari padanan kepada data ZEUS.*

Substruktur kuark juga diperlihat jika didapati wujud keadaan kuark teruja, dan carian kuark teruja boleh dibuat menerusi proses dalam Gambarajah 10. Data dari ZEUS, seperti dalam Gambarajah 11, memberikan zon pengecualian di antara 40 GeV dan 169 GeV kepada jisim kuark teruja [5].

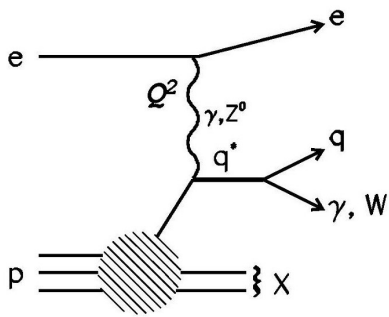

*Gambarajah 10. Penghasilan kuark teruja.*

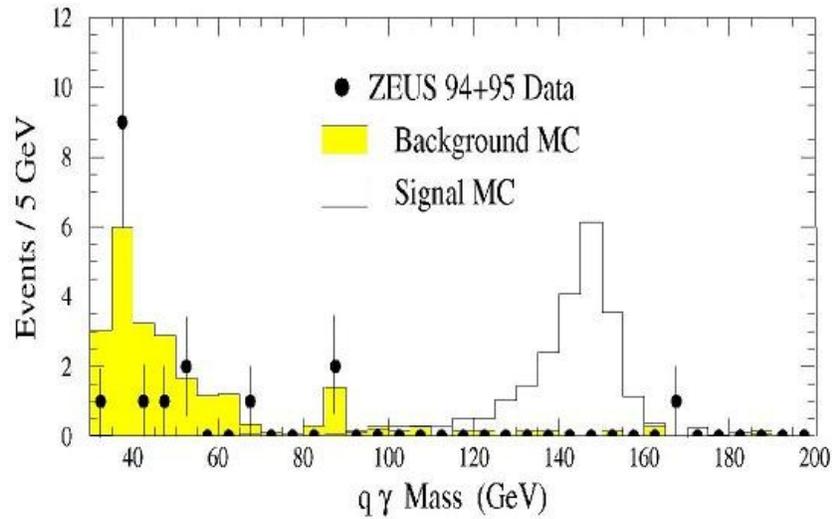

*Gambarajah 11. Carian kuark teruja menerusi resonans qγ dalam data ZEUS.*

**Keratan rentas dan fungsi struktur**

Dalam model kuark-parton, apabila hanya kuark valens diambilkira, keratan rentas diberi dengan mudahnya,

$$F_2(x) = x \sum_i e_i^2 f_i(x)$$

yang tak bergantungkan $Q^2$ (penskalaan Bjorken). Suatu 'kejutan' oleh HERA ialah pencanggahan penskalaan ini, terutama pada pecahan momentum kecil, seperti dipaparkan dalam Gambarajah 12 dan 13. Banyak parton kelihatan pada $x$ rendah, terutama pada $Q^2$ tinggi. Di sini pemerihalan QCD diperlukan untuk menerangkan banyak parton (maya) yang dilihat pada momentum rendah, yang selaras dengan kebebasan asimptot.

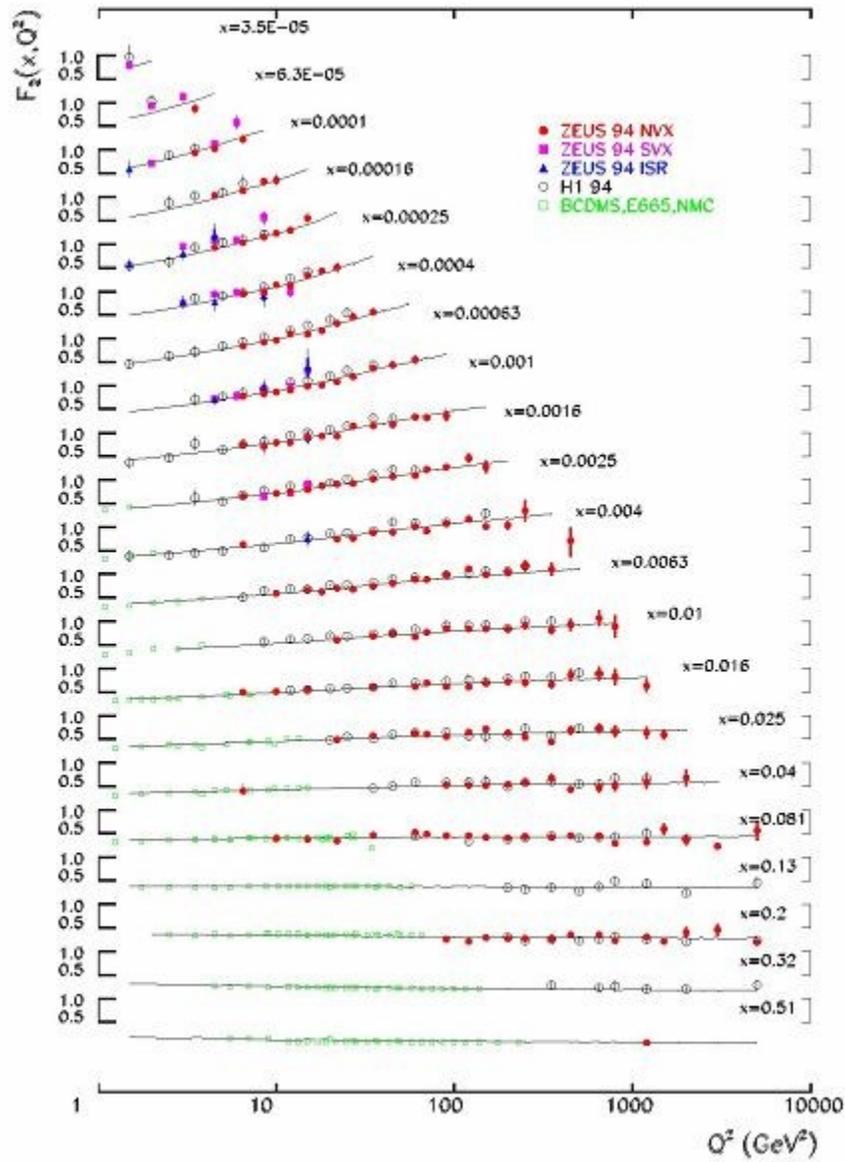

Gambarajah 12. Keratan rentas DIS terhadap $Q^2$ untuk nilai-nilai x berlainan. Dari [6].

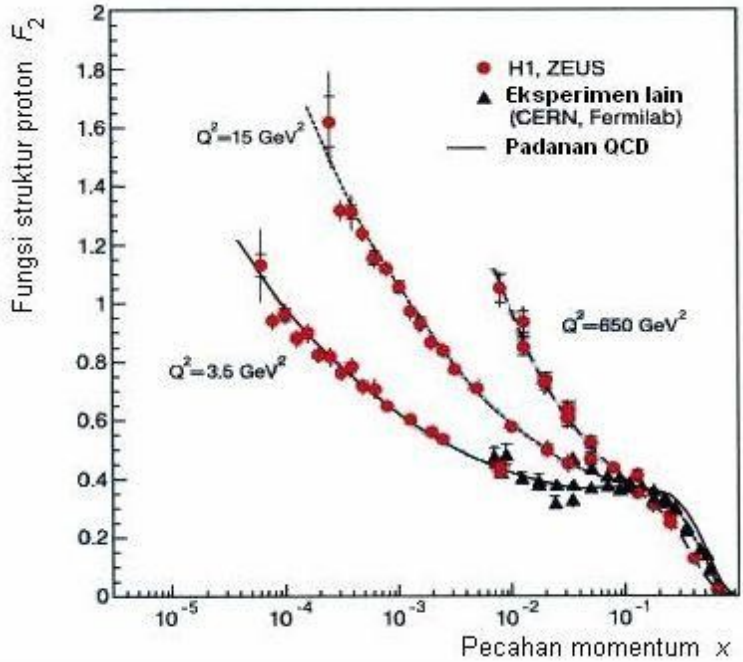

*Gambarajah 13. Keratan rentas DIS terhadap x untuk nilai-nilai $Q^2$ berlainan. Dari [7].*

Pengeluaran perisa berat pula, kerana bukan merupakan perisa valens, adalah peka kepada kandungan gluon di dalam proton (Gambarajah 14). Keratan rentasnya juga memperlihatkan telatah yang sama (Gambarajah 15), dengan itu menyarankan gejala yang sama bagi gluon dan kuark.

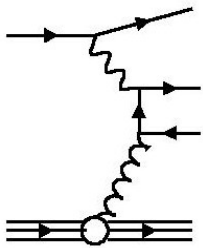

*Gambarajah 14. Penghasilan perisa berat dari pelakuran foton-gluon.*

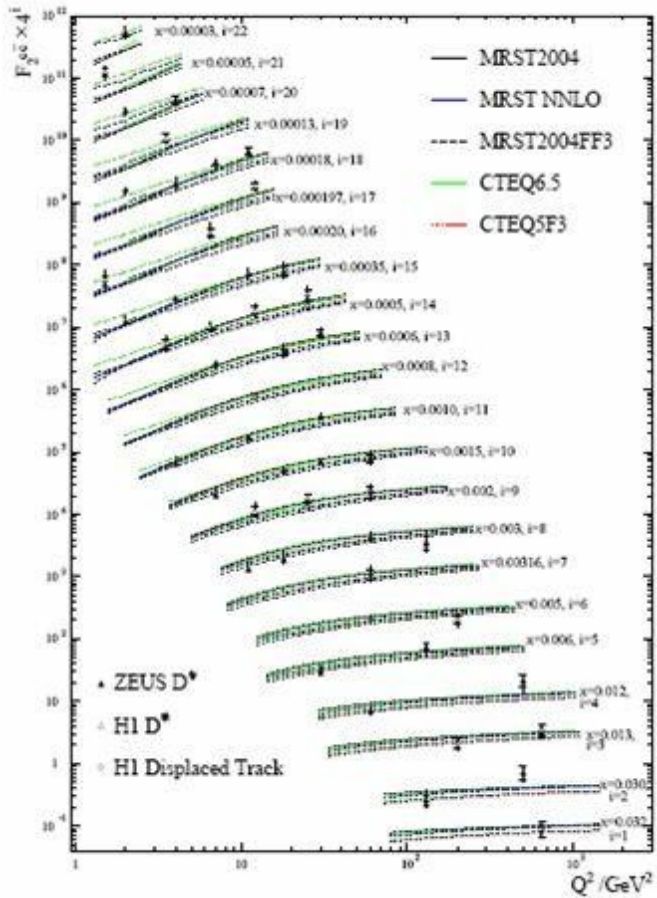

*Gambarajah 15. Keratan rentas bagi keluaran pesona. Dari [8].*

**Proses belauan**

Satu lagi 'kejutan' dari HERA ialah banyaknya sumbangan proses belauan. Satu tanda proses belauan ialah ada jurang kecepatan yang besar di antara sisa proton dan hasil yang lain. Kecepatan,

$$\eta = \ln\left(\frac{E - p_L}{E + p_L}\right)$$

mengukur 'kelajuan membujur', dan jurang kecepatan mencerminkan tiada aliran warna, seperti diterangkan dalam Gambarajah 16. Ini bermakna foton menghentam sesuatu yang mempunyai nombor kuantum vakum. Entiti ini dinamakan pomeron, IP.

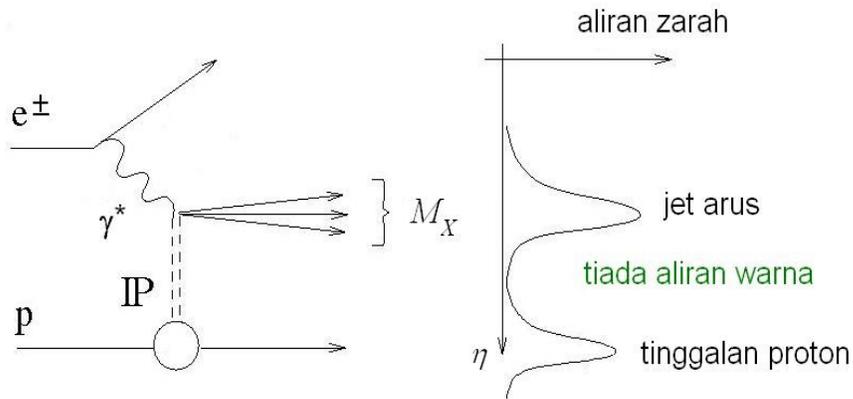

*Gambarajah 16. Peristiwa dengan jurang kecepatan besar.*

Peristiwa-peristiwa belauan (Gambarajah 17), mempunyai parameter tambahan, kemayaan pomeron,

$$t = (p - p')^2,$$

jisim berkesan jet arus, $M_X$, dan pecahan momentum parton dalam pomeron,

$$\beta = \frac{Q^2}{2(p-p') \cdot q} = \frac{x}{x_{IP}} = \frac{Q^2}{Q^2 + M_X^2 - t}$$

Seperti ditunjukkan dalam Gambarajah 18, proses belauan memberi sumbangan hampir 10 % kepada DIS, bagi $Q^2$ rendah.

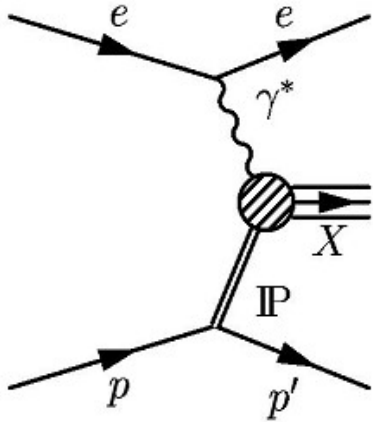

*Gambarajah 17. Proses belauan.*

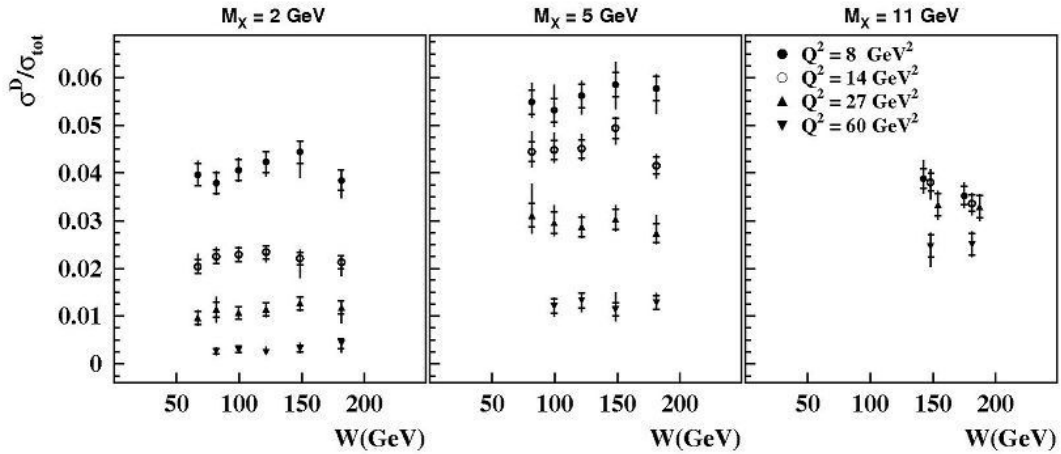

*Gambarajah 18. Sumbangan proses belauan kepada DIS. Dari [6].*

Persoalan yang dibawa oleh proses-proses belauan ini ialah perlunya pefahaman tentangnya, khasnya apa itu pomeron dari segi pemerihalan QCD. Kemungkinannya ia adalah struktur gluon ynag tertentu, yang menghasilkan nombor kuantum vakum tadi. Fungsi struktur pomeron dalam proton telah dikaji (Gambarajah 19), yang juga berbau pencanggahan penskalaan.

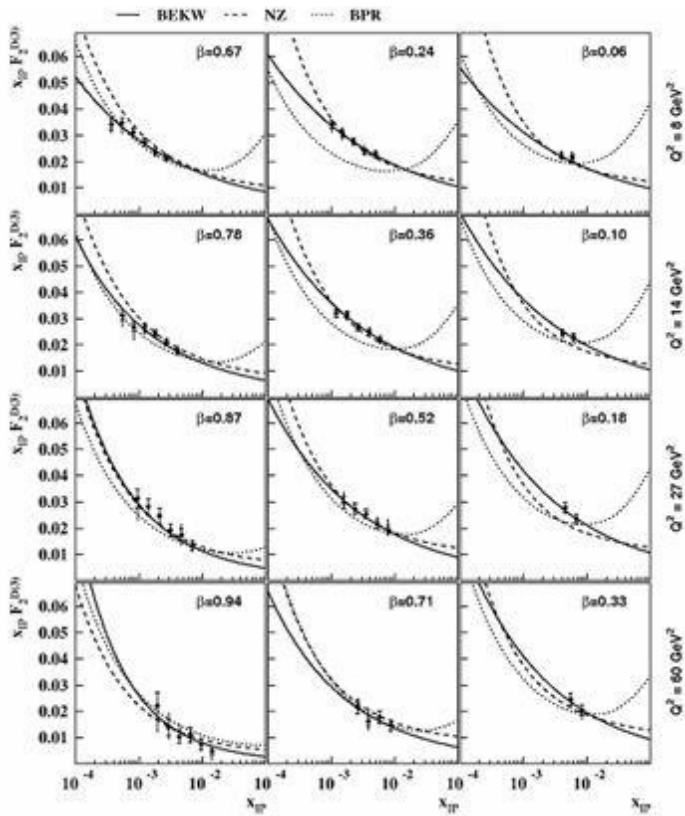

*Gambarajah 19. Fungsi struktur pomeron dalam proton berbanding pelbagai model. Dari [6].*

**Gambaran proton**

Pendugaan elektron bertenaga tinggi (di bawah "mikroskop elektron" attoskala) ke atas proton memberikan kita gambaran yang kaya – banyak parton-parton 'halus' (Gambarajah 20) dan wujudnya struktur pomeron. Banyak lagi sumbangan kepada gambaran proton dari misalnya, penghasilan eksklusif meson vektor dalam DIS, yang kita tidak bincangkan di sini, tetapi memberi pengayaan yang selari.

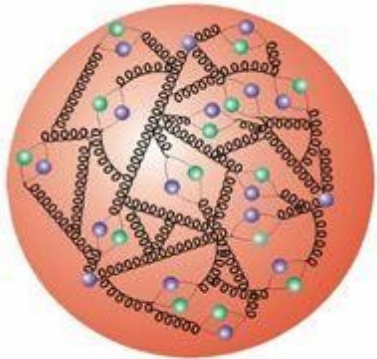

*Gambarajah 20. Gambaran proton di bawah mikroskop attoskala. Diambil dari [7].*

**Penutup**

Kita telah lihat usaha memahami struktur QCD hadron menerusi perlanggaran elektron-proton pada tenaga tinggi. Banyak lagi fizik QCD dan bukan-QCD yang menarik dalam eksperimen sebegini yang tidak disentuh di sini.

Eksperimen di perbatasan ilmu sebegini juga menuntut teknologi berkait yang terkehadapan, seperti yang diperihalkan berkenaan pembacaan data. Ia juga membabitkan kerja berpakatan yang antarabangsa yang membuka jalan kepada kesejahteraan sejagat. Ini agak ikonan kepada Pulau Duyong ini.

*Pecah ruyung mencari manisan*
  *Masamnya pauh ditelan jamu*
*Kota Duyong jadi labuhan*
  *Berlayar jauh ke lautan ilmu*

## Rujukan